\documentclass[AMA,Times1COL]{WileyNJDv5}

\articletype{Literature Review}%

\received{Date Month Year}
\revised{Date Month Year}
\accepted{Date Month Year}
\journal{Journal}
\volume{00}
\copyyear{2023}
\startpage{1}

\begin{document}

\title{A Review of Password-less User Authentication Schemes.}

\author[1]{Tunde Oduguwa}

\author[1]{Abdullahi Arabo}


\titlemark{A REVIEW OF PASSWORD-LESS USER AUTHENTICATION SCHEMES}

\address[]{\orgdiv{Department of Computer Science and Creative Technologies}, \orgname{University of the West of England}, \orgaddress{\state{Bristol}, \country{England}}}

\corres{Corresponding author: Tunde Oduguwa; University of the West of England, Bristol, United Kingdom. \email{oduguwaoa@yahoo.ca}}

\abstract[Abstract]{Since the demise of the password was predicted in 2004, different attempts in industry and academia have been made to create an alternative for the use of passwords in authentication, without compromising on security and user experience. This review examines password-less authentication schemes that have been proposed since after the death knell was placed on passwords in 2004. We start with a brief discussion of the requirements of authentication systems and then identify various password-less authentication proposals till date. We then evaluate the truly password-less and practical schemes using a framework that examines authentication credentials based on their impact on user experience, overall security, and ease of deployment. The findings of this review observe a difficulty in balancing security with a user experience compared to that of passwords in new password-less schemes, providing the opportunity for new applied research to leverage existing knowledge and combine technologies and techniques in innovative ways that can address this imbalance.}


\maketitle

\renewcommand\thefootnote{}
\footnotetext{\textbf{Abbreviations:} OWASP, Open Worldwide Application Security Project; NIST, National Institute of Standards and Technology; FI, Federated Identity; CAPTCHA, Completely Automatic Public Turing Test to Tell Computer and Human Apart; ECC, Elliptic Curve Cryptography; BB, Behavioural biometrics; KD, Keystroke Dynamics; IMEI, International Mobile Equipment Identity; CA, Certificate Authority; FIDO2, Fast Identity Online; RFID, Radio Frequency Identification; SIM, Subscriber Identity Module .}

\renewcommand\thefootnote{\fnsymbol{footnote}}
\setcounter{footnote}{1}

\section{Introduction}\label{sec1}

According to the Open Worldwide Application Security Project (OWASP), "Authentication is the process of verifying that an individual, entity or website is whom it claims to be"\cite{AuthenticationOWASPCheat}. National Institute of Standards and Technology (NIST) further expands that authentication involves a party (the claimant) successfully demonstrating "possession and control of one or more authenticators (previously registered) to another party (the verifier) through a protocol" \cite{NISTSpecialPublication}. As passwords have been the major authenticator for a long time, implementation of authentication systems has revolved around policies that describe acceptable password creation and management. While these policies were created for user security, they have come at the expense of security, necessitating alternative authentication paradigms not involving passwords. In theory, a proper authentication system should balance user security with a wonderful experience, but the reality observed in new authentication schemes is that there are still difficulties in combining technologies in ways that will replace the password scalably. The aim of this review then is to identify the approaches that have been used to implement password-less alternatives. This will guide research of a new scheme that mitigates the challenges faced in building a scalable and secure password-less authentication system having a great user experience.

We start the review by exploring existing research into various password-less schemes, in chronological order. We then provide a definition for schemes which we consider to be truly password-less, and evaluate the truly password-less schemes using the framework proposed by Bonneau et. al\cite{bonneauQuestReplacePasswords2012}. Their framework suggests minimum criteria required for a properly designed authentication system.

\section{Password-less Authentication Timeline}\label{sec2}

Password-less authentication is a system of granting access to some resource, in a scalable way, without the use of shared secrets like passwords. Since about '300 billion'\cite{campbellPuttingPassePasswords2020} passwords are used and currently shared, leading to millions of them being stolen daily\cite{campbellPuttingPassePasswords2020}, the drastic need for a more secure alternative has prompted continuous research into ways that will eliminate passwords as we currently know them. 

\subsection{2005 - 2010}
The earliest observed proposal to replace passwords was a concept known as Federated Identity - FI\cite{shimFederatedIdentityManagement2005}. 

Since users normally require access to multiple digital resources simultaneously and needing separate passwords, FI was created to manage this situation. It eliminates the need for multiple passwords by delegating authentication responsibility to a trusted party, after initial password authentication to that party. In this scenario, the trusted party maintains a network of participating services to which a user can be authenticated by proxy using their original credentials. While this reduces the burden of password management on the user, it is dependent on a central master password, ignoring all the other failings of such an arrangement. 

While FI was memory-based, Chetty et. al (2007)\cite{chettyPasswordLessSecurity2023}, implemented an attributes-based system where authentication depended on a combination of multiple biometric attributes of a user like voice, facial features, lips, etc. The rationale is that while single factor biometrics like fingerprints, iris, gestures, etc uniquely identifies an individual, it is challenging to consistently achieve the degree of accuracy required for authentication purposes\cite{bhattacharyyaBiometricAuthenticationReview2009}. In effect, a fusion of biometrics was better suited to compensate for any observed shortcomings of one or the other individual biometric attribute of the user. 

During the same period, Shirali-Shahreza and Shirali-Shahreza (2007)\cite{mohammadshirali-shahrezaPasswordlessLoginSystem2007} took a device-centric approach by designing a system that captures the embedded IMEI number (a unique ID attached to every mobile phone). During login, a user is presented with a "CAPTCHA" - a test that is easy for humans but difficult for bots"\cite{vonahnCAPTCHAUsingHard2003}. Once the user solves the CAPTCHA challenge, the user's response and IMEI are combined to authenticate the user. Although Shirali-Shahreza and Shirali-Shahreza refer to this solution as a password-less login, we argue that this is not the case as the scheme does not satisfy the conditions described by OWASP and NIST for authentication. Further, their implementation suggests anyone in possession of the device can access resources behind the login wall.
\subsection{2010 - 2020}

Previous research has identified an inverse relationship between password strength and memorability\cite{gehringerChoosingPasswordsSecurity2002}, prompting studies into a visual approach for user authentication. “Graphical passwords”\cite{biddleGraphicalPasswordsLearning2012} leverage the ability of users to easily recall images and/or patterns, rather than text. The reality is that visual schemes dependent on recognition of images\cite{biddleGraphicalPasswordsLearning2012}, while facilitating password-less authentication are subject to the same vulnerabilities as text-based passwords; they require the user accurately recollecting the exact context of the image. 

At another end of the spectrum from graphical passwords, weaving credentials into stories has also been considered to replace the password\cite{wernerCognitiveApproachesPassword2012}. Depending on the implementation mechanics, this proposes to be a highly secure system as it depends on the users’ unique history/experience/memories. A drawback of this system is the observed usability: while users are the sole author of their stories, it can be challenging to reproduce the exact context used during the initial registration process. We experienced this exact phenomenon during our research for this review; a subject that built their password based on a historical event could not access an account for several hours as they struggled to remember the exact context of their own story that was used to create the authentication credentials. It is vulnerable to a form of unconscious self-created denial of service. 

Veering off in a different direction, cryptography, a process that enables secure communication over insecure channels\cite{goldwasserLectureNotesCryptography2008} was the central theme of the system put forward by\cite{zhuLoxinSolutionPasswordless2014}. Asymmetric encryption, involving the use of public and private keys was used to identify the user. However, the system also involves two additional third parties that complete the authentication process: a certificate authority and a push-based message service. While this system also depends on a device password as in the case of Shirali-Shahreza and Shirali-Shahreza, (2007)\cite{mohammadshirali-shahrezaPasswordlessLoginSystem2007} without which it will be available to anyone in possession of the device, the attack surface is large as there are several external points from which failure could disrupt the system. 

The CAPTCHA and mobile device IMEI were again implemented in another password-less attempt by Kalra, (2016)\cite{kalraNOVELPASSWORDLESSAUTHENTICATION2016}. However, asymmetric encryption used was the differentiator from the similar earlier work by Shirali-Shahreza and Shirali-Shahreza, (2007) \cite{mohammadshirali-shahrezaPasswordlessLoginSystem2007}. Elliptic curve cryptography was used in this revised system to create a digital signature from the CAPTCHA input supplied by the user before it is sent along with the device IMEI to a server for the verification. Regardless of the use of encryption in this case, the system is dependent on device password or pin and subject to the same vulnerability as a password; in the case where the user has not set any access credential on the device, anyone in possession of the device can access the user's account. 

The use of biometrics has long been considered as an alternative for user authentication. While one form of biometrics depends on the attributes or physical features of a user like fingerprint or voice, another form explores the user's behaviour and is referred to as behavioural biometrics - BB. It deals with human behaviours observed during normal interaction with digital devices. Behaviours like typing patterns, mouse movement and motion, hand signature patterns, etc\cite{khareBehaviouralBiometricsCognitive2013}, are some examples. 

Jadhav et. al,(2017)\cite{jadhavBiometricAuthenticationUsing2017} investigated how keystroke dynamics -KD (a form of behavioural biometrics) could offer a better alternative for authentication over passwords. They described it as "the automated method of identifying or confirming the identity of an individual based on the manner and rhythm of typing on a keyboard"\cite{jadhavBiometricAuthenticationUsing2017}, which is unique to each individual and offers several advantages. However, factors such as fatigue, drug influence, injury, etc., can negatively impact the operation of a KD-based system, necessitating continued research into the concept.

To offset shortcomings observed in single behaviour-based authentication systems, Tanwar et. al, (2019)\cite{tanwarApproachEnsureSecurity2019} proposed a biometric system that fused user features with behaviour (as a second factor) i.e., voice and behaviour like typing pattern or hand gesture. According to research by Meng et. al, (2020)\cite{mengActiveVoiceAuthentication2020}, "The voice is unique because the construction of the articulatory organs and its use that generate and modulate sounds are uniquely configured for each individual and is naturally embedded in the voice's nature." Due to this attribute of the voice, there are potential benefits in the proposal by Tanwar et. al, (2019)\cite{tanwarApproachEnsureSecurity2019} when combined with other behavioural biometrics\cite{abozaidMultimodalBiometricScheme2019} of the user, although little was provided in terms of implementation details and requires further research if it is to be a realistic solution.

\subsection{2020 - Onwards}

Most of the behavioural biometrics schemes we identified up to the year 2020 involve some conscious action by the user. However, the human brain has also been observed to produce (unconscious) unique identification data in the form of electrical signals that can be captured using non-invasive means\cite{poulosNeuralNetworkBased1999}. Based on this, brain-based authentication was proposed by Kopito et. al, (2021)\cite{kopitoBrainbasedAuthenticationScalable2021} as a secure method of user authentication. In this scheme, the user's brain signals are captured and processed using machine learning models to extract features that are stored for use during authentication. One limitation of this authentication system is the initial enrolment process; this activity will be carried out in a controlled environment with qualified personnel managing the procedure in clinical fashion. Alternatively, users would be required to wear special headgear under strict guidelines. In view of this, further research work is necessary if brain signals are viable as a replacement for passwords. 

Another biometric feature that is unique is the iris (a part of the human eye), such that "even the two eyes of an individual and identical twins possess uncorrelated patterns"\cite{matinHumanIrisBiometric2016}, and was the basis of the authentication system proposed by Mohammed and Ahmed, (2022)\cite{mohammedIRISRECOGNITIONTECHNOLOGY2022}. While iris-based systems have been recorded to have the highest degree of accuracy of all biometrics, implementing it as a password alternative requires extensive research. Take the enrolment process for instance: " a specialised camera, typically within three feet of the subject, uses an infrared imager to illuminate the eye and capture an extremely high-resolution photograph" \cite{kopitoBrainbasedAuthenticationScalable2021}. Unless current devices are upgraded with the capability for such functionality or future devices engineered for this purpose, enrolment will have to be conducted in conditions like that of brain-signal authentication, making it infeasible for mass adoption. 

Continued research into biometric authentication was reflected in the work of Progonov et. al, (2022)\cite{progonovBehaviorbasedUserAuthentication2022} when they proposed the fusion of multiple behaviours. User behaviours such as typing, swiping, app usage, grip patterns, device gesture and eye motion tracking are collected and a profile is built with this data, to authenticate the user. One major advantage is the context-awareness of this model: if the system identifies behaviour inconsistent with that of the user, there is a chance that a third-party might have hijacked the user's session - the session is ended until the user is re-identified. A flaw Progonov et. al's proposal, however, is that an enrolment procedure is not discussed. Authentication based on such a model requires some form of initial registration - it is assumed that this scheme will be an extended functionality (or second factor) for an existing authentication system, which might be password based. 

The recurring theme from the schemes reviewed suggests that a secure authentication system with great user experience appears to be a holy grail for security researchers, influencing the decision of technologies being combined in a myriad of ways to achieve the desired result. Conners et. al, (2022)\cite{connersLetAuthenticateAutomated2022} proposed the use of certificates issued by a CA (Certificate Authority) as an improvement to an existing password-less scheme - FIDO2\cite{bicakciFIDO2PasswordlessAuthentication2022},\cite{ghorbanilyastaniFIDO2KingslayerUser2020}. 
Considering that FIDO2 is dependent on hardware tokens and does not offer account recovery options besides registering at least a second token as backup, making it an expensive proposition, Conners et. al, (2022)\cite{connersLetAuthenticateAutomated2022} in their work describes how certificates can address this gap, allowing users to potentially achieve password-less nirvana by combining a hardware token and a software authenticator, where the token is responsible for managing authenticators rather than authentication itself. However, it is crucial that users register a backup token for account recovery purposes; the difference between this and FIDO2 is that only a single backup token is needed, rather than multiple tokens in the case of FIDO2. 

A major weakness of password-based systems is that anybody in possession of the password can access the resource, as it is not "tied" to the user. Huang et. al, (2022)\cite{huangRFIDAuthenticationSystem2022} proposed to create a link between a user and their password using RFID such that even if someone else was in possession of that same password, they would not be authenticated as the user. Per Huang et. al, while setting their password, user's body gives off bioelectrical signals which are captured and mapped to the password, such that the password does not have to be secret. Like earlier proposed brain-based and iris authentication systems, special hardware is required to capture the user's body signals during enrolment. Further, the authors describe that "during the registration phase, users need to provide multiple sets of data (a minimum of 20 sets) as training data"\cite{huangRFIDAuthenticationSystem2022}, which suggests that the system might be infeasible to replace the password on a large scale without additional research.

Mobile operating systems have been observed to implement password-less authentication schemes where the user authenticates by drawing patterns through joining dots on a grid of the device. Andriotis et. al, (2022)\cite{andriotisBuDashUniversalDynamic2023} built upon this concept and proposed an improved system based on a grid made up of different shapes like circles, stars, triangles, etc instead of static dots. Further, the shapes change as the user draws the unlock pattern, making the system more secure. While this scheme is designed to work only on mobile devices (Android based), we argue that the scheme is also low on usability, like strong passwords of great length that users will find challenging to keep in memory and write in a text file. Another mobile technology that considered to replace the password is SIM. Rao and Bakas, (2023)\cite{raoAuthenticatingMobileUsers2023} proposed an expanded use of the SIM beyond authenticating users to mobile networks, considering that the technology is also capable of authentication to other digital resources and currently has spare capacity that can be deployed for this purpose. Before this can be deployed for full-scale adoption as a password alternative however, Rao and Bakas describe a requirement for "some extensions to the current mobile standards to facilitate the new features" needed by the scheme, amongst others.

While this timeline is not an exhaustive list of every attempt made to replace the password since 2004, it reflects the diversity of approach by different researchers, the difficulty faced when attempting to balance security with a great user experience, as well as provides a strong foundation for the future research combining various technologies to create an acceptable alternative to passwords.

\section{Evaluation of Password-less Authentication Schemes}\label{sec3}

Following their review of "two decades worth of proposals"\cite{bonneauQuestReplacePasswords2012}  by different research groups attempting to create password-less alternatives, Bonneau et. al's, (2012) framework based on "a broad set of twenty-five benefits expected of an ideal scheme"\cite{bonneauQuestReplacePasswords2012}  revealed the difficulty inherent in achieving password-less nirvana. This review will also evaluate the schemes identified earlier using the same framework to identify gaps and subsequently form the basis of future research that will ensure a secure and scalable password-less system is achieved.

Their framework defines "25 properties framed as a diverse set of benefits" as a set of objective criteria which a scheme proposing to replace the password should adhere to across three categories namely usability, security, and deployability. 

Table 1 describes the items that makes up these categories. 

\begin{center}
\begin{table*}[!h]%
\caption{Bonneau et. al's (2012) framework elements.\cite{bonneauQuestReplacePasswords2012}\label{tab1}}
\begin{tabular*}{\textwidth}{@{\extracolsep\fill}lllll@{}}
\toprule
\textbf{BENEFIT} & \textbf{DESCRIPTION}   \\
\midrule
Memorywise-Effortless & Users don't have to remember any secrets    \\
Scalable-for-Users & Using the scheme for many accounts doesn't increase user burden     \\
Nothing-to-Carry & No need to carry any additional physical objects     \\
Physically-Effortless & No physical effort beyond simple actions like pressing buttons    \\
Easy-to-Learn & Users can easily figure out how to use the scheme   \\
Efficient-to-Use & Authentication time is reasonably short   \\
Infrequent-Errors & Low error rates for legitimate users   \\
Easy-Recovery-from-Loss & Users can conveniently regain authentication ability if credentials are lost   \\
Accessible & Usable by people with disabilities   \\
Negligible-Cost-per-User & Very low cost of deployment per user   \\
Server-Compatible & Compatible with existing server-side authentication systems   \\
Browser-Compatible & Works with existing standard browsers, no plugins required   \\
Mature & Tested outside of research, with standards and implementations  \\
Non-Proprietary & No royalty fees or restrictions on use  \\
Resilient-to-Physical-Observation & Resistant to observation attacks like shoulder surfing  \\
Resilient-to-Targeted-Impersonation & Resistant to impersonation using personal knowledge \\
Resilient-to-Throttled-Guessing & Resistant to online guessing attacks \\
Resilient-to-Unthrottled-Guessing & Resistant to offline brute force attacks \\
Resilient-to-Internal-Observation & Resistant to malware key-loggers or eavesdropping \\
Resilient-to-Leaks-from-Other-Verifiers & Leaks from one verifier don't help attackers impersonate the user elsewhere \\
Resilient-to-Phishing & Resilient to phishing attempts  \\
Resilient-to-Theft & Physical theft of authentication object doesn't enable impersonation  \\
No-Trusted-Third-Party & No trusted third party that could compromise security \\
Requiring-Explicit-Consent & Authentication can't happen without explicit user consent  \\
Unlinkable & Colluding verifiers can't determine if the same user is authenticating to both \\
\bottomrule
\end{tabular*}
\begin{tablenotes}
\end{tablenotes}
\end{table*}
\end{center}

Having identified the framework elements, we define a scheme as truly password-less if users are not expected to keep the credentials in memory. As an example, the scheme by Shim, Bhalla and Pendyala, (2005) known as federated identity\cite{shimFederatedIdentityManagement2005} is dependent on a central password created at the service that provides the user's authentication credentials to other registered services that have implemented the scheme, without which it would not be possible for a user to use the scheme. In this case, we do not consider this as true password-less and will not be evaluating it using Bonneau et. al's, (2012)\cite{bonneauQuestReplacePasswords2012} framework.  

It is crucial to mention that technological advancements have occurred after the framework was proposed and will be factored in our evaluation. This means that benefits previously low on some elements at the time of the framework might now be considered high if a scheme leverages advancements that have made them possible. For example, at the time of the proposal, fingerprint biometrics was rated low on the 'Infrequent-Errors' benefit due to an inefficient experience during the authentication procedure. Since that time however, improvements in research and industry have made fingerprint biometrics more efficient and used on a wider scale, guiding us to rate schemes involving fingerprint biometrics high on the benefits relating to user experience. The following section identifies research application gaps still existing.

\subsection{Implementation Gaps Identified in Literature}
For a scheme to replace the password practically, all the components on which the scheme is based should already exist or be in the process of development by the team or other interested parties. Short of this, the goal to replace the password would remain in the realm of theoretical research. Considering that this review is conducted with the aim of applying the findings to create a working password-less authentication scheme, we believe that only schemes involving the use behavioural biometrics like keystroke dynamics\cite{jadhavBiometricAuthenticationUsing2017} and certificates-based\cite{connersLetAuthenticateAutomated2022} schemes meet our criteria. This is because the components needed to accomplish these schemes into actual password-less schemes are current available for implementation, while others are still theoretical at best.

A key challenge with keystroke dynamics-based password-less schemes is that there are frequent errors with authenticating a user based on their typing patterns. According to Jadhav et. al, (2017)\cite{jadhavBiometricAuthenticationUsing2017}, the accuracy of the scheme can be impacted by "injury, fatigue, drug influence or distraction", leading to false negatives and positives. Further, if a database of user typing models is hacked, the "Easy-Recovery-from-Loss" benefit becomes non-existent; since a user cannot realistically change their typing pattern, they will need to be moved to another system of authentication.

While the certificates-based scheme by Conners et. al, (2022)\cite{connersLetAuthenticateAutomated2022} offers significant benefits, it is low on the 'Nothing-to-Carry' and 'Negligible-Cost-per-User' options as users incur extra costs in purchasing at least 2 hardware tokens; a main and back-up token. Since there is negligible financial cost to using a password, only users wishing to protect exceedingly high value resources will consider this scheme. Our review therefore suggests that research into innovative ways of combining proven technologies to build a password alternative should be is still necessary to be undertaken in future works.

\section{Conclusion}
In this paper, password-less authentication schemes have been reviewed and analysed. Having identified the difficulty in balancing security, deployability with usability, we believe that further applied research is needed to address these persistent gaps if passwords will be phased out. Considering that previous research has already provided a good foundation, our future work will consider the use of symmetric and asymmetric cryptography, biometrics and steganography, in a password-less scheme that can address the gaps identified by this review.  

\bibliography{references}

\end{document}